# Wafer-scale integration of single nanodiamonds via electrostatic-trapping

Jixiang Jing[1]†, Yicheng Wang[1]†, Zhuoran Wang[2], Yumeng Luo[3], Linjie Ma[1], Tongtong Zhang[1], Chunlin Song[4], Jiangyu Li[4], Kwai Hei Li[3], Dong-Keun Ki[5], Ji Tae Kim[6], Zhiqin Chu[1]*

**Affiliations:**

[1]Department of Electrical and Electronic Engineering, The University of Hong Kong; Hong Kong, China.

[2]Department of Mechanical Engineering, The University of Hong Kong; Hong Kong, China.

[3]School of Microelectronics, Southern University of Science and Technology; Shenzhen 518055, China.

[4]Department of Materials Science and Engineering, Southern University of Science and Technology; Shenzhen 518055, China.

[5]Department of Physics and HK Institute of Quantum Science & Technology, The University of Hong Kong; Hong Kong, China.

[6]Department of Mechanical Engineering, Korea Advanced Institute of Science and Technology (KAIST); Science Town, Daejeon 34141, Republic of Korea.

†These authors contributed equally to this work.

*Corresponding author: zqchu@eee.hku.hk (Zhiqin Chu)

**Abstract:**

Nanodiamonds (NDs) are key materials for building nanoscale quantum sensing, imaging and communication devices. Scalable configuration of single NDs on heterogeneous platforms, forming photonic quantum source arrays, will be an essential solution towards realizing next-generation practical and industrial quantum devices. However, NDs are challenging to manipulate because their size, shape and surface chemistry vary substantially. Here, we show a simple method based on electrostatic-trapping to rapidly and reliably pattern single ND arrays on arbitrary substrates at scale. Our method, which uses carefully engineered microscale hole templates and electrostatic force, captures single NDs across 8-inch wafers with 82.5% yields within 5 min. Systematic experimental and theoretical studies show the number of deposited NDs primarily depends on the diameter of the hole trap. The method is compatible with mature CMOS technologies, enabling the mass production of scalable and integrable quantum devices. This advancement is expected to accelerate the commercialization and industrial adoption of ND-based technologies.

**Main:**

Diamond's ability to host quantum defects – most notably nitrogen-vacancy (NV) centers – makes it an exceptional material for advancing quantum technologies [1,2,3,4]. Nanodiamonds (ND), in particular, are ideal for quantum metrology, sensing and imaging [5,6,7,8] because they are small (typically 1-200 nm) [9], hard [10], chemically stable [11] and biocompatible [12]. As heterogenous integration becomes more popular in advanced microelectronics and photonic integrated systems (PICs) [13, 14, 15, 16, 17], the ability to add arrays of single NDs into these systems on-demand and at scale will be an essential step towards realizing next-generation monolithic quantum devices that are multifunctional and programmable [18, 19, 20].

Manipulating NDs, however, is challenging. Their size, shape and surface chemistry are inherently heterogenous, unlike spherical silicon dioxide ($SiO_2$), polystyrene (PS) and metallic nanoparticles [21, 22, 23]. Template-assisted methods [24, 25, 26], electrostatic self-assembly [27, 28, 29, 30], three-dimensional (3D) printing [31, 32] and atomic force microscope (AFM) tip-based transfer [33, 34] are currently used to pattern ordered single NDs. However, these methods are tedious and can only yield a few single ND across millimeter-scale areas. Moreover, producing miniature hole and disk templates for confining NDs require time-consuming and expensive electron beam lithography setups, which are incompatible with complementary metal-oxide semiconductor (CMOS) technology.

Here, we report a simple CMOS-compatible method based on electrostatic-trapping to reliably integrate single ND arrays onto arbitrary substrates at scale, filling a long-term gap between laboratory and industrial scale production of ND-based quantum devices. Inspired by template-assisted patterning, where NDs are confined within holes or disks, and concepts in optical tweezers, where optical field gradient force is used to confine miniature objects within the narrow waist of a laser beam [35, 36, 37], we constructed an electrostatic field force gradient inside a template hole to electrostatically anchor single-NDs at the center of the hole. To ensure the electrostatic field force is distributed within the narrow space inside the template hole, we introduced positive charges on the template surface and maintained negatively charged template sidewalls. The resulting spatial electrostatic field distribution gradient allows only electronegative NDs, which are typically functionalized with -COOH groups, to pass through the hole and anchor to the positively charged template surface at the bottom of the hole (**Fig. 1a**).

For our experiments, we used NDs with an average diameter of 150 nm (**Fig. 1b** and **Supplementary Fig. 1**). The NDs were processed via salt-assisted air oxidation to remove surface-adsorbed impurities, such as ultrasmall NDs and disordered carbons, and dispersed in an aqueous solution (**Methods**). To obtain the trapping template, we treated a silicon (Si) substrate in oxygen plasma for 10 min to increase its hydrophilicity (**Supplementary Fig. 2**) and chemically modified the resulting hydroxylated surface with positively charged amino ($NH_3^+$) functional groups by immersing the substrate in 5% 3-aminopropyltriethoxysilane (APTES) solution for 3 hours (**Supplementary Fig. 3** and **Methods**). The corresponding chemical reaction are displayed as follows:

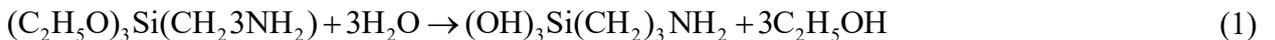

$$(C_2H_5O)_3Si(CH_23NH_2) + 3H_2O \rightarrow (OH)_3Si(CH_2)_3NH_2 + 3C_2H_5OH \quad (1)$$

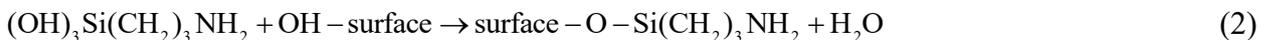

$$(OH)_3Si(CH_2)_3NH_2 + OH-surface \rightarrow surface-O-Si(CH_2)_3NH_2 + H_2O \quad (2)$$

A layer of negatively charged photoresist is then spin-coated onto the modified substrate and circular holes are patterned using photolithography (**Fig. 1c**). Trapping involves covering the holes with aqueous ND (0.1 mg/mL) and letting it sit at room temperature for 5 minutes (**Supplementary Fig. 4** and **Movie 1**).

Single ND arrays formed after the 5-min incubation (**Fig. 1b**), and the remaining aqueous ND is collected with a pipette for reuse. Immersing the substrate in acetone removes the template in one minute. ND concentrations that are either too low or too high caused insufficient trapping, indicating that the number of NDs matter and overly high numbers of NDs can interfere with the trapping electrostatic field (**Supplementary Fig. 5**). Trapping single NDs in the holes depends on the incubation time and hole diameter (**Extended Data Fig. 1**). For a 3.6 μm diameter hole, a single ND is trapped after 5 min and remained single for 3 hours before additional NDs populated the hole. Longer incubation times can damage the photoresist and compromise the internal electrostatic field. Larger holes (≥ 6 μm) trapped multiple NDs and became saturated over time.

Our method is scalable and tunable. Large-area ND arrays observable with naked eyes can be easily obtained on an 8-inch Si wafer (**Fig. 1d**). Dark-field optical images of arrays from different regions of the Si wafer show the ND patterns are highly uniform and clean (**Fig. 1e**). Timely collection of the aqueous ND with the pipette is key. Residues form on the patterned arrays when the ND solution is allowed to self-evaporate (**Supplementary Note. 1** and **Extended Data Fig. 2**). Additionally, the small ND arrays and large mark 'HKU' on the wafer illustrate the flexibility of our method, enabling the ND patterning from micro to macro scale.

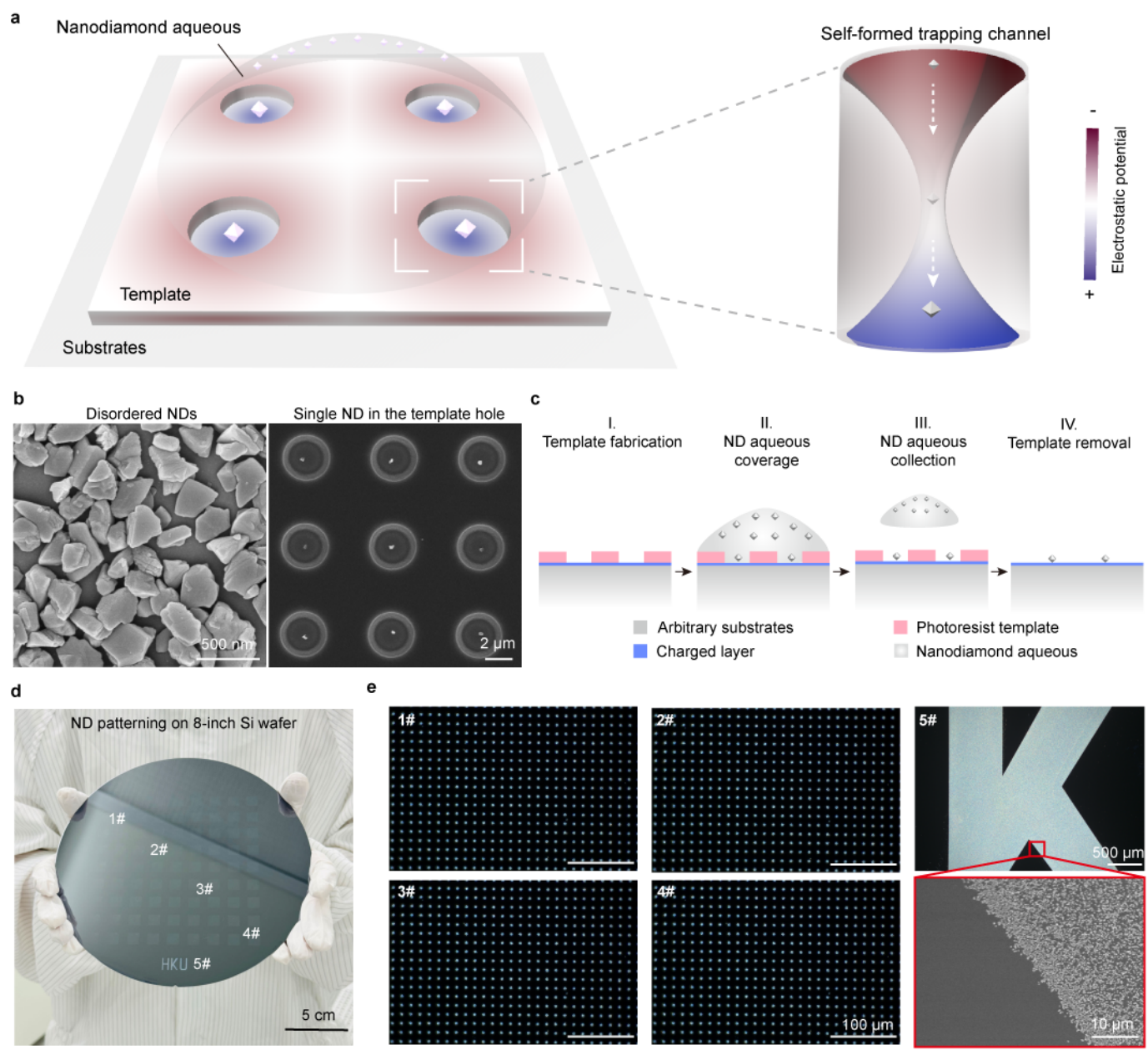

**Fig. 1 Electrostatic trapping as a scalable method to produce clean and uniform ND arrays on CMOS-compatible substrates.**
**a**, Schematic illustrating our electrostatic-trapping method, which involves covering a template containing circular holes with aqueous ND. Blow-up diagram on right shows the electrostatic potential distributed inside the hole presents an hourglass shape that only allows single NDs to pass through the hole. **b**, Scanning electron microscope (SEM) images of NDs before (left) and after (right) ordering them into arrays of single NDs using our method. **c**, Schematic diagram showing the template hole fabrication and trapping process.

Photoresist template is created using photolithography (I). Aqueous ND is delivered and collected using a pipette (II, III). Template is removed using acetone (IV). **d**, Photograph of an 8-inch silicon wafer containing multiple ND arrays (white squares). Numbered areas are magnified in (**e**). **e**, Dark-field optical images show ND arrays from numbered areas (#1-4) in (**d**) are uniform and clean. Dark-field image (top) of the letter 'K' (#5) and the magnified SEM image (red box) show NDs distributed inside the letter 'K' formed a clean boundary.

We show ND trapping through electrostatic absorption is feasible in our setup. Kelvin probe force microscopy (KPFM) measurements show that amino-modified holes displayed significantly higher surface electrostatic potential (~112 mV) than bare, unmodified ones (~40 mV) (**Fig. 2a**). Notably, surface electrostatic potential saturates easily for substrates that were incubated for longer than 3 hours in APTES (**Extended Data Fig. 3**). The photoresist template performed an electrostatic potential of -13 mV relative to the bare substrate. Additionally, NDs subjected to salt-assisted air oxidation displayed high zeta potential with an average value of -35 mV (**Fig. 2b**). These negatively charged NDs were trapped only in positively charged amino-modified holes (**Fig. 2c**), regardless of hole diameter (**Supplementary Fig. 6**). No NDs were seen in all sizes of bare, unmodified holes. Furthermore, when other ions that could break down the trapping electrostatic field such as $Na^+$ and $Cl^-$ ions were introduced into the aqueous ND solution, no trapping was observed (**Supplementary Fig. 7**). These results together strongly suggest that ND trapping occurred through electrostatic absorption.

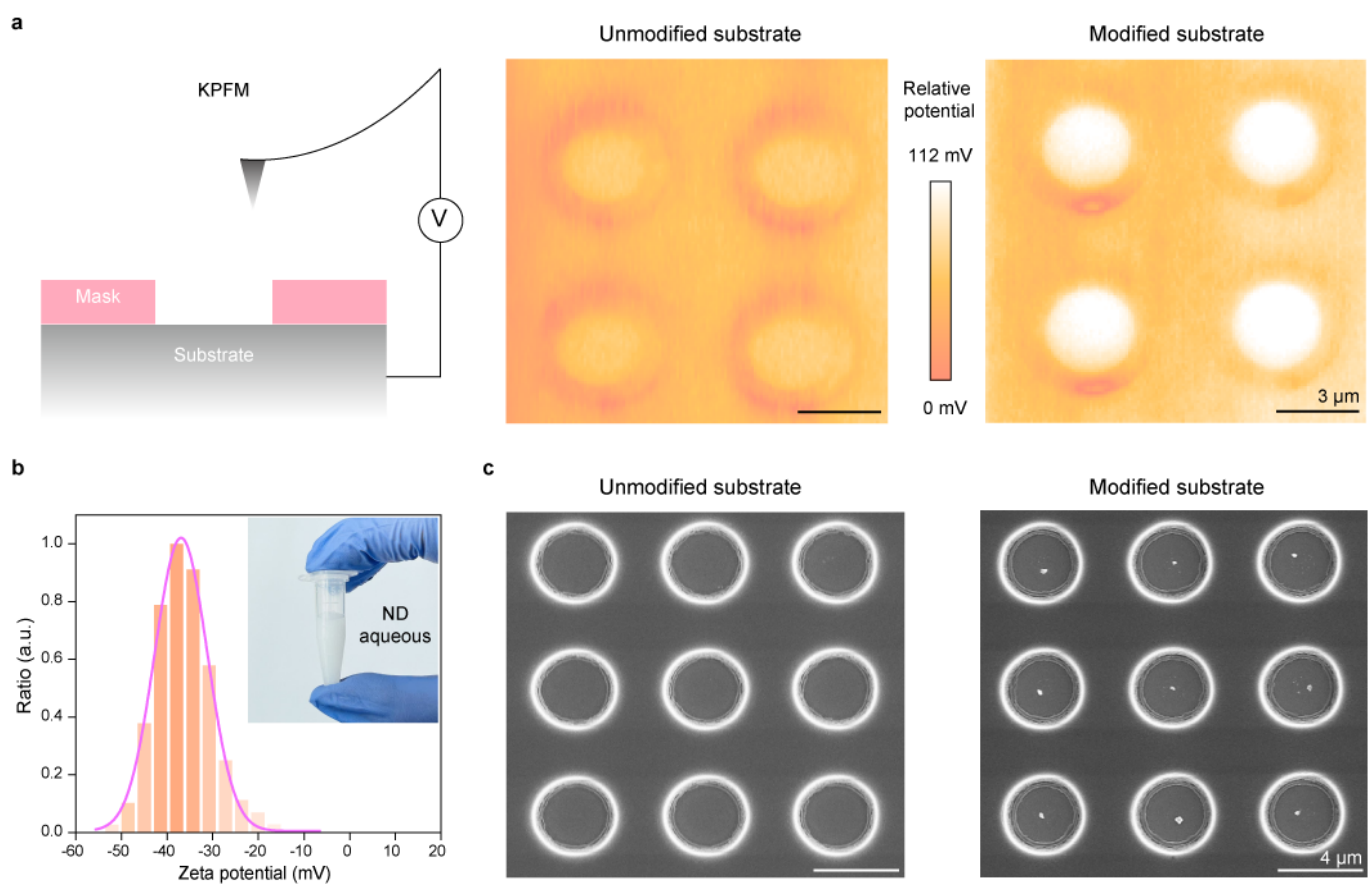

**Fig. 2 ND trapping in template hole is driven by electrostatic absorption.**
**a**, KPFM measurement of a template hole (left). Amino-modified holes (right) displayed a significantly higher surface electric potential than bare, unmodified holes (middle). **b**, Aqueous NDs used in our experiments display an average zeta potential of -35 mV. Inset: photograph of measured NDs. **c**, SEM images show single NDs are trapped in holes modified with a charged layer (right), but not bare, unmodified holes (left).

We further verified that the hourglass shape trapping method proposed in Fig. 1a is not influenced by poor aqueous infiltration in the template hole. Our experiments show water completely infiltrated 3-µm wide holes after 10 min (**Supplementary Fig. 8**). To determine the trapping conditions and configuration in the template hole (**Fig. 3a**), we used finite element method (COMSOL Multiphysics) and mean-field Poisson-Boltzmann (PB) theory [38, 39] to calculate the spatial distribution of electrostatic potential in the template hole (**Methods**). Applying appropriate boundary conditions and solving the non-linear PB equations in the model, we find that the electrostatic potential $\psi$ inside the hole is non-uniform, decaying either from top to bottom or from the edge to the center of the hole (**Fig. 3b**). This condition naturally causes a negatively charged particle to move towards a location with lower potential, that is, to the center and bottom of the hole as observed in our experiments. Based on standard electric potential threshold values for triggering the movement of charged particles, we further identified the center of the hourglass shape as the region most likely to trap a ND (**Extended Data Fig. 4**). These theoretical results are consistent with the experimental results and the proposed trapping model in Fig. 1a.

Because hole diameter affects the number of trapped NDs, we also calculated the effective trapping region for holes of different diameters (**Extended Data Fig. 4**). By quantitatively comparing the minimal horizontal width of the effective trapping volume marked in Fig. 3b, we found that the trapping width enlarges linearly with hole diameter, $D$ (**Fig. 3c**), but not hole depth, $h$ (**Fig. 3d**). For a 3 µm diameter hole, the trapping width at different $h$ was consistently around 0.4 µm. In agreement with these calculations, our experimental results show that trapping single ND is more sensitive to $D$ than $h$ (**Fig. 3e** and **Supplementary Fig. 9**). Consequently, the number and size of trapped NDs can be easily tailored through $D$. Our trapping method is universal and can be used to capture other types of particles, including silica, polystyrene and gold nanoparticles (**Supplementary Fig. 10**).

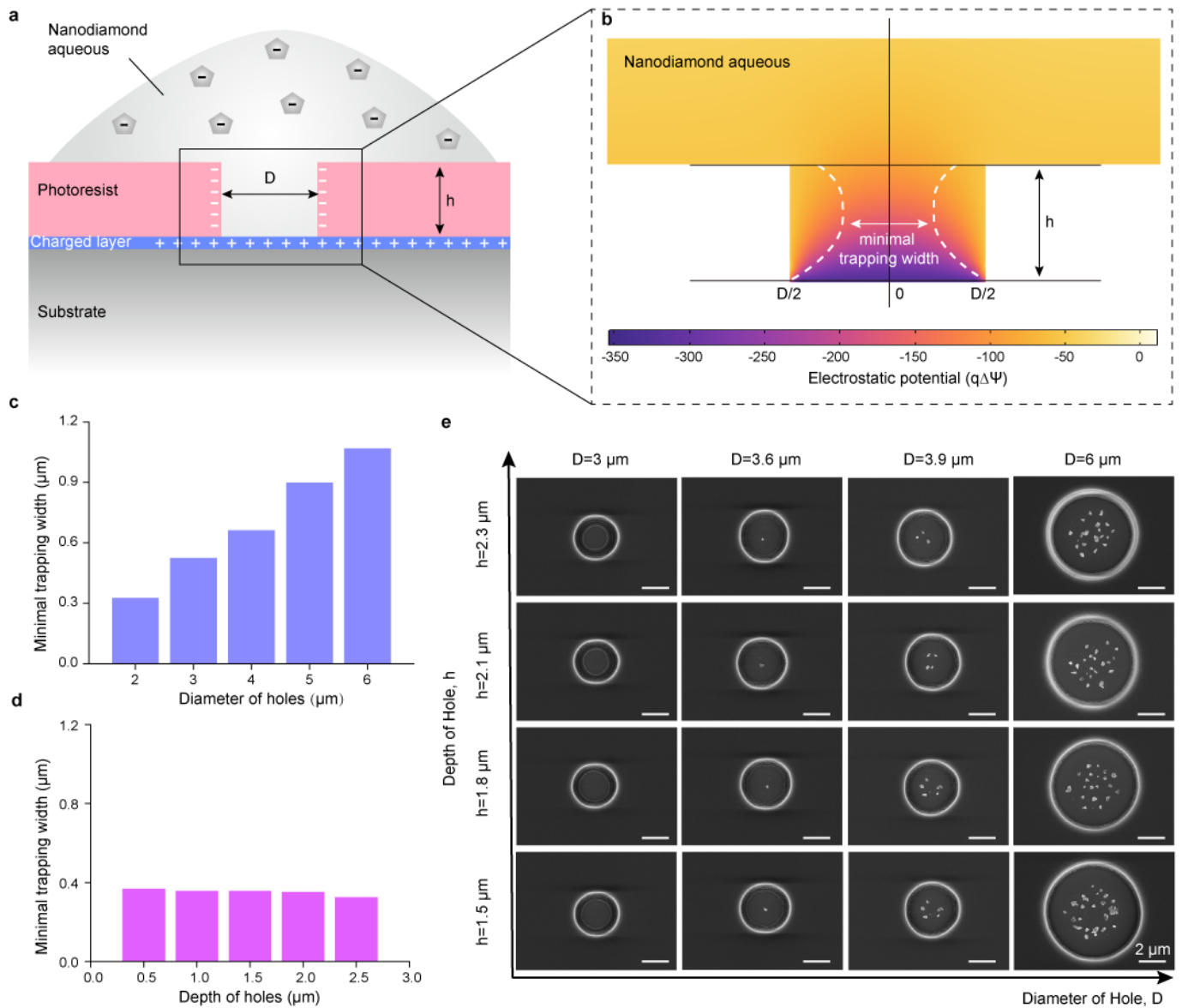

**Fig. 3 ND trapping depends on hole diameter, but not hole depth.**
**a**, Schematic illustrating the surface charge on the substrate and photoresist template that make up a patterned hole. **b**, Distribution of calculated electrostatic potential in the template hole. Dotted white lines outline the threshold potential region. Anywhere within the outlined region will effectively trigger ND movement. White arrow represents the minimal width of the effective trapping region. **c**, Minimal trapping width increased linearly with hole diameter, $D$. **d**, Minimal trapping width for a 3 μm diameter hole is stable across different hole depths, $h$. **e**, SEM images of trapped NDs in hole templates of different diameters (column) and depths (rows). Hole diameter ranged from 3 μm to 6 μm while hole depth varied from 1.5 μm to 2.3 μm.

We repeated each ND trapping experiment 5 times and confirmed the method is highly reproducible even for a 25 × 25 array (**Figs. 4a-d**). Examining the distributed position of trapped NDs, we found that most NDs were concentrated within 500 nm from the hole's center (**Fig. 4e** and **Extended Data Fig. 5**). This concentrated area is consistent with the calculated minimal trapping width in Fig. 3 and shows that the trapping is highly accurate. Counting the number of single ND in the entire array after template removal, we found that up to 82.5% of the array had a single ND (**Figs. 4c, d, f** and **Extended Data Fig. 6**). This single ND yield and 8-inch wafer patterning area is highest amongst other single ND placement methods [24, 27, 31] (**Fig. 4g**). In principle, our approach can be extended to the largest commercially available Si wafer (12-inch), and the single ND yield can be further improved through more precise photolithography setups.

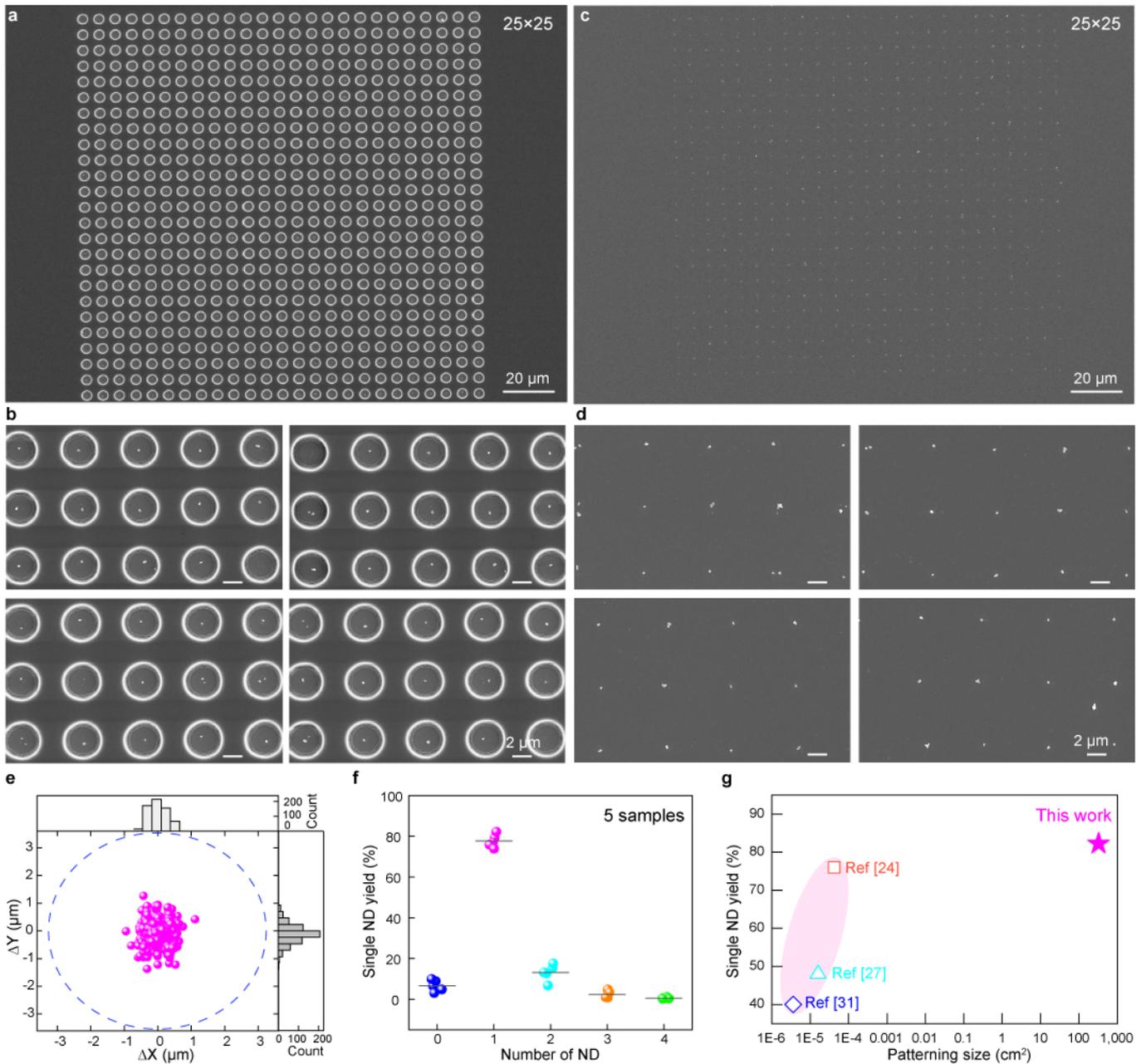

**Fig. 4 Electrostatic-trapping method for single ND placement is highly reproducible.**
**a** and **c**, SEM images showing a 25 × 25 ND array before (**a**) and after (**b**) template removal. **b** and **d**, Enlarged SEM images of four 5 × 3 ND arrays randomly selected from (**a**) and (**c**), respectively. Single ND arrays are consistently obtained. The NDs remained in place even after template removal. **e**, Statistics of the relative positions of NDs within the template holes of a 25 × 25 array. Most NDs were trapped within 500 nm from the hole's center. **f**, Statistical analysis across five 25 × 25 array samples show 82.5% template holes contained a single ND. **g**, Compared to other methods (Ref. 24, 27, 31) used to obtain single ND array, our electrostatic-trapping method achieved the largest patterning area and highest single ND yield.

The electrostatic-trapping method is compatible with hybrid platforms and current CMOS technology. We fabricated scalable heterogenous platforms such as silicon waveguides, gallium nitride (GaN) pillars, and gold microwave antennas through mature CMOS technology and used our electrostatic-trapping method to integrate single ND arrays (**Fig. 5a** and **Supplementary Note. 2**). After template removal, the trapped NDs remained in position (**Fig. 5b** and **5c**). Compatibility with CMOS technology means that our method can be broadly applied to a host of other heterogenous platforms and devices.

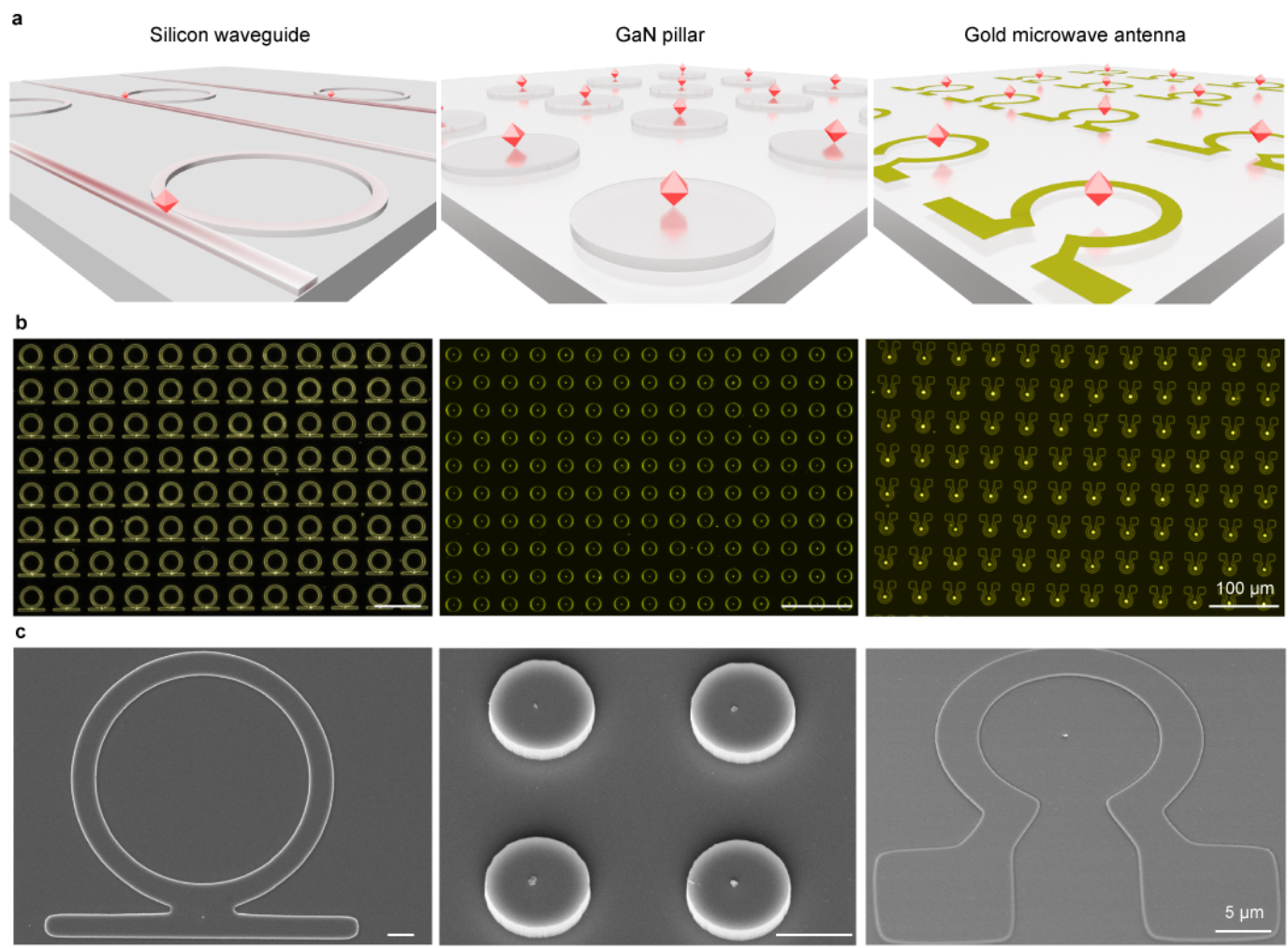

**Fig. 5 Electrostatic-trapping method is compatible with hybrid platforms and CMOS technology.**
**a-c**, Schematic (**a**), dark-field optical images (**b**) and high-magnification SEM images (**c**) show single NDs are integrated at precise locations on a silicon waveguide (left), GaN pillar (middle) and gold microwave antenna (right) array.

**Conclusion**

In conclusion, we have developed an electrostatic-trapping method to rapidly integrate single ND array with CMOS-compatible platforms at scale. Using standard photolithography and electrostatic field force, we realized reliable trapping of single ND array across large areas (8-inch wafer scale) with high single ND yields (82.5%). Unlike current strategies, our method is simple, highly reproducible, CMOS-compatible and scalable to any commercially available platform. The experiment and simulation results show that trapping is made possible by the non-uniform distribution of electrostatic potential inside the trapping template hole. An hourglass-like trapping channel within the hole controls the passage of single ND. We show the method can be used to place single ND arrays on heterogenous platforms fabricated using standard CMOS technology. The ability to reliably order single NDs at scale on any platform through such a simple method is expected to impact the development and commercialization of various ND-based quantum sensing, imaging, communication and computing technologies.

## Methods

ND processing via salt-assisted air oxidation
NDs with an average size of 150 nm (HPHT, PolyQolor, China) were first mixed with sodium chloride (NaCl, 99.5%, Sigma-Aldrich) at a ratio of 1:5, followed by heating at 500 °C for 5 h in air. Next, the mixtures of ND and NaCl were dispersed in deionized (DI) water and sonicated for 1 h, and then NDs were purified with DI water for 3 times by centrifugation (first: 1000 rcf, 5 min; second: 3000 rcf, 5 min; third: 8000 rcf, 10 min). Finally, the purified NDs were re-dispersed in DI water and sonicated for 10 min to obtain well-dispersed ND suspension for further use.

Template preparation
The employed silicon (Si) wafer was first cleaning in ultrasound bath in acetone, isopropyl alcohol (IPA), DI water for 10 min, respectively, followed by drying with nitrogen gas. Then, the clean Si wafer was treated by oxygen plasma for 10 min to increase surface hydrophilicity. After that, the Si wafer was immersed in the solution of ethanol and silane agent (3-Aminopropyltriethoxysilane) with a ratio of 1:30 for 3 hours at room temperature.

After completion of surface modification on Si wafer, the standard micro/nano fabrication process was utilized to generate the template pattern. First, the photoresist AZ5214 was spin-coated on Si wafer at a rate of 4500 r/min for 50 s, followed by post-baking at 90 °C for 60 s. Then, the template pattern was defined by photolithography at a UV dose of 300 mJ/cm$^2$ and then developed for 60 s. The preparation of trapping template was therefore well achieved.

ND trapping processes
After preparation of ND aqueous and trapping template, the ND aqueous was diluted to 0.1 mg/mL and sonicated for 20 min to realize sufficient dispersion. Then, using a pipette to control a certain volume of ND solution and drop on the template, in which the solution should completely cover the template pattern. Next, keeping the coverage for 5 min at room temperature, followed by collection of ND solution from the template by the pipette. To further remove the template, the wafer after completion of ND trapping should be immersed in acetone solution for 1 min, followed by drying with nitrogen gas.

Calculating the spatial distribution of electrostatic potential in template holes
According to the geometrically induced electrostatic trapping mechanism in fluids, the corresponding model in our case was built based on the practical measurements in Fig. 2. As shown in Fig. 3a, the bottom layer represents a Si substrate, while the holes on its top are formed by a layer of photoresist. When in contact with DI water, the surface of the holes can ionize and dissociate chemical groups, resulting in electrostatic potentials on both the Si substrate and the photoresist walls, referred to as $\varphi_{bottom}$ and $\varphi_{wall}$, respectively. Using the mean-field Poisson-Boltzmann (PB) theory [38, 39], we can calculate the electrostatic potential free energy of the ND particles in this system. By analyzing the distribution of the electrostatic potential, we can understand why ND particles tend to deposit at the center of the holes and the relationship between the deposition distribution and the size of the holes.

By setting appropriate boundary conditions, we can solve the nonlinear PB equation and obtain the spatial distribution of the electrostatic potential by using COMSOL Multiphysics [38, 39, 40]. Near an isolated ionized surface, the electrostatic potential $\psi$ decays exponentially with distance, following the relationship [41]:

$$\psi(z) = \psi_0 \exp(-\kappa z) \qquad (3)$$

Among them $\psi_0$ is the surface value, $\kappa^{-1}$ represents the Debye length, in electrolyte solution [42],

$$\kappa^{-1} = \sqrt{\frac{2c_0 e^2}{\varepsilon \varepsilon_0 k_B T}} \tag{4}$$

$c_0$ represents the salt concentration in the electrolyte, $\varepsilon \varepsilon_0$ is the dielectric constant of the medium, and $e$ is the elementary charge. In a cylindrical hole-shaped hole with a diameter of $2r$, the electrostatic potential decays monotonically from the circular hole wall and reaches a minimum at the center when the fluid fills the entire hole. We define a dimensionless PB equation and assume that the electrostatic potential can be expressed as:

$$\nabla^2 \psi = (\kappa r)^2 \sinh(\psi) \tag{5}$$

where $r$ represents the radius of the trap and $\kappa^{-1}$ represents the inverse Debye length based on the condition of the solution. We assume that the interior of nanodiamonds is a uniform dielectric environment, with a dielectric constant of $\varepsilon_p$, namely, $\nabla^2 \psi = 0$. Because the electrostatic potential of wall is constant, we set the Dirichlet boundary condition:

$$\psi_s = \psi_w \tag{6}$$

By measuring surface potential using KPFM, the surface potential of the photoresist in solution with pH=7 $\psi_{wall} \approx$ -13mV, while surface potential of silicon in DI water is positive, $\psi_{bottom} \approx$ +72mV (after silanization processing). Setting the dimensionless electrostatic potential $u$ to facilitate calculation, which is also one of the parameters required for simulation results:

$$u = \frac{e\psi}{k_B T} \tag{7}$$

For the particle properties, measurements have shown that the surface potential is approximately -35mV, and the diameter is around 150 nm. Based on previous statistical measurements and data analysis [43, 44], the particle charge $q$ is estimated to be approximately -120e. The simulation results are shown in Fig. 3b, where the cross-sectional electrostatic potential is represented as $q\psi$ in units of $k_B T$.

It is generally believed that when $q\Delta\psi$ is approximately 10 $k_B T$, the particle can acquire sufficient trapping force to remain in a relatively stable position within the hole. From Fig. 3b, it can be observed that at the center of the same horizontal cross-section, the electrostatic potential is the lowest. The dashed lines represent the positions where $q\Delta\psi$ is equal to 10 $k_B T$ at each height, and the regions between them can be referred to as "traps" for capturing particles (**Fig. 3b** and **Extended Data Fig. 4**). When the hole diameter is 2 μm, the trap region is the smallest, which facilitates the convergence and deposition of particles towards the center of the bottom. As the hole diameter increases, the trap gradually widens, leading to the gradual dispersion of particles at the bottom and an increase in the deposition radius. Since the particles carry charges of the same sign and repel each other, when the trap diameter is small, fewer particles can fit inside, resulting in a reduced number of deposited nanoparticles. Conversely, when the trap is wide enough, more particles can pass through it, leading to an increased number of particles at the bottom. This explains why a smaller hole diameter corresponds to fewer deposited particles, which are more prone to converge towards the center, while a larger hole diameter leads to more deposited particles and a wider deposition area.

**Acknowledgement**

We thank A.L. Chun of Science Storylab for valuable discussions and for critically reading and editing the manuscript.

**Funding:**

Z.C. acknowledges the financial support from the National Natural Science Foundation of China (NSFC) and the Research Grants Council (RGC) of the Hong Kong Joint Research Scheme (Project No. N_HKU750/23), RGC Theme-based Research Scheme (Project No. T45-701/22R), and the Health@InnoHK program of the Innovation and Technology Commission of the Hong Kong SAR Government. J.T.K. acknowledges support from the National Research Foundation of Korea (NRF) grant funded by the Korea government (MSIT) (No. RS-2025-00556379, RS-2024-00407084). D.K.K. acknowledges the financial support from the RGC of Hong Kong SAR under the scheme of Area of Excellence (AoE/P-701/20). J.L. and C.S. acknowledge the financial support from National Key Research and Development Program of China (2022YFF0706100).

**Author contributions:**

J.J. and Z.C. conceived the idea. Z.C. supervised the project. J.J. led the experimental investigation with assistance from Y.W., Z.W., Y.L., L.M., T.Z., C.S.. J.J. wrote the manuscript with input from all authors. Z.C., J.T.K., D.K.K., K.H.L., J.L. discussed the results and commented on the manuscript.

**Competing interests:**

Z.C. and J.J. are inventors on a Chinese Invention Patent related to this work (No. 202510992503.X) entitled "Process for formation of patterned functional nanoparticles on a substrate". The authors declare no competing financial interests.

**Data and materials availability:**

All data are reported in the main text and the supplementary materials.

**Extended Data Figures**

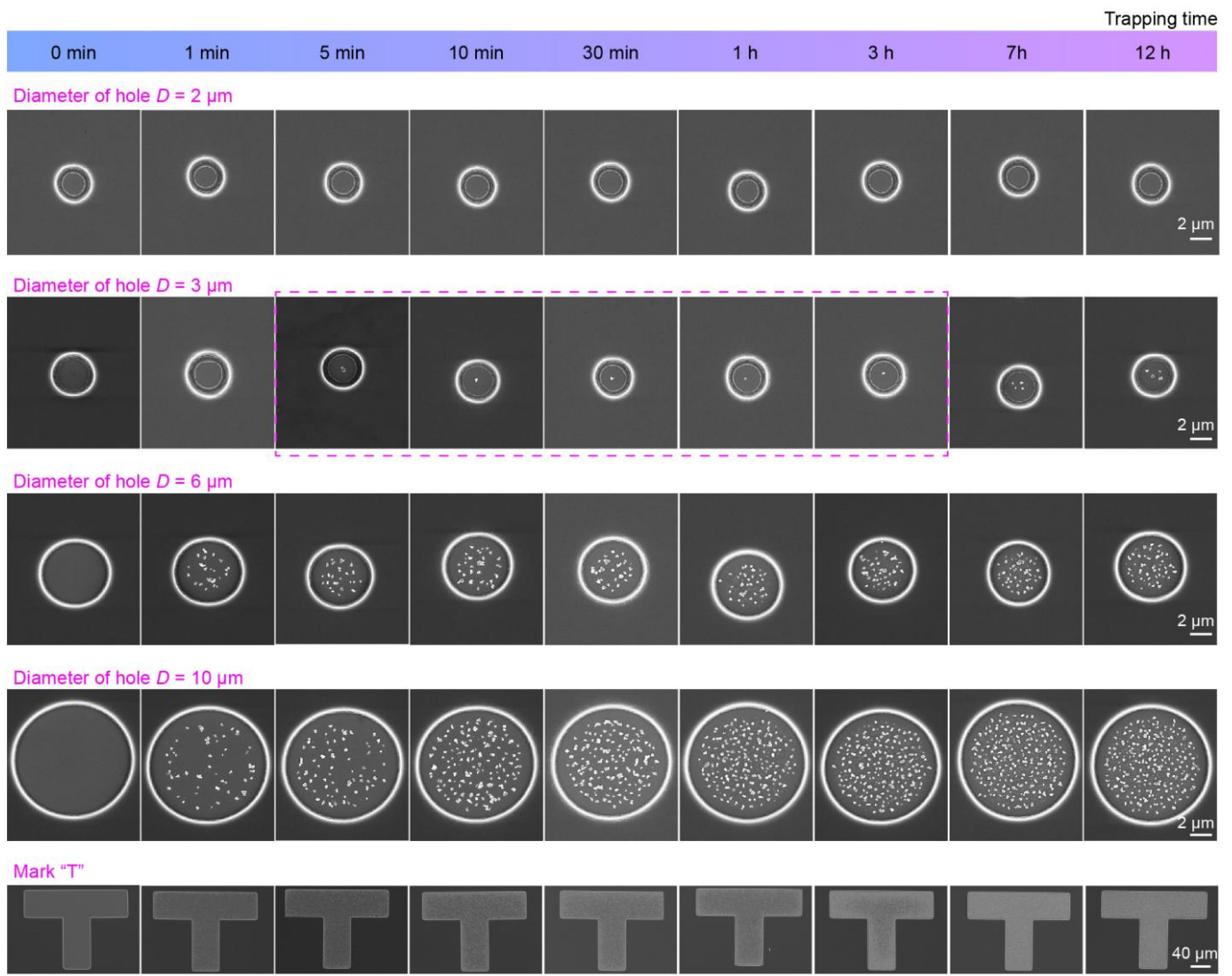

**Extended Data Fig. 1**

The trapping performance for different-sized template holes in various trapping times. For the smaller holes with a diameter of 2 µm, there was no trapped NDs all the time. Starting from the size of 3 µm (actual size of 3.6 µm, the difference was induced by photolithography error), there was occurring single ND after 5 min and lasting for 3 hours. For a long-term trapping (> 7 h), there were multiples NDs in the hole that mainly attribute to the damage of photoresist soaked for a long time, in which the internal electric filed was broken down. For the larger traps (size > 6 µm), the number of NDs was gradually increasing at first 10 min and then saturated at the later hours.

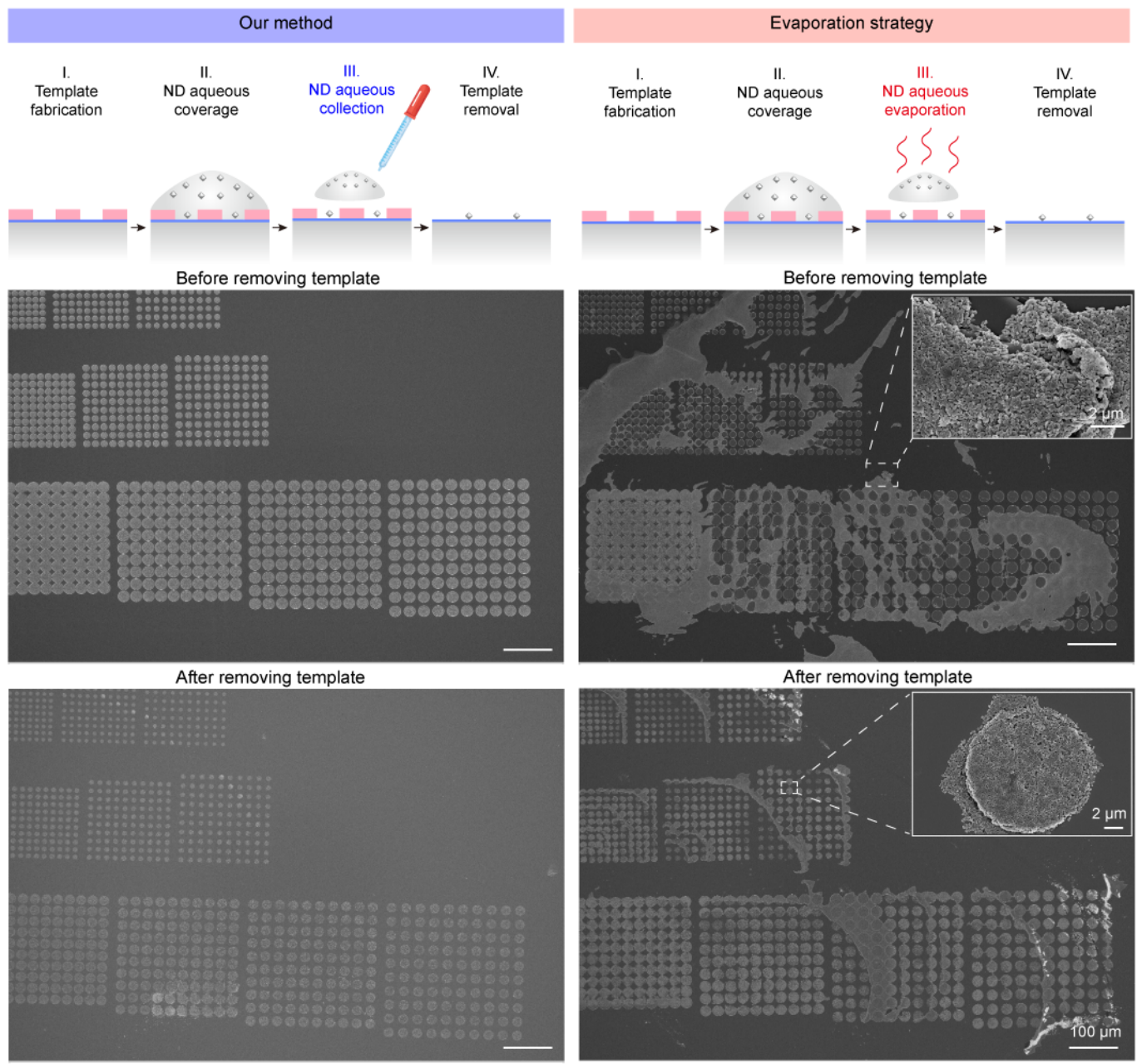

**Extended Data Fig. 2**

Comparison of our method and evaporation strategy. The top layer indicates the fabrication procedure of these two ways, in which the ND aqueous was timely collected after trapping in our method in contrast to self-evaporation in another one. The resulting SEM images show that there are many residues on the pattern for the evaporation strategy, while the trapping pattern is cleaning by our method.

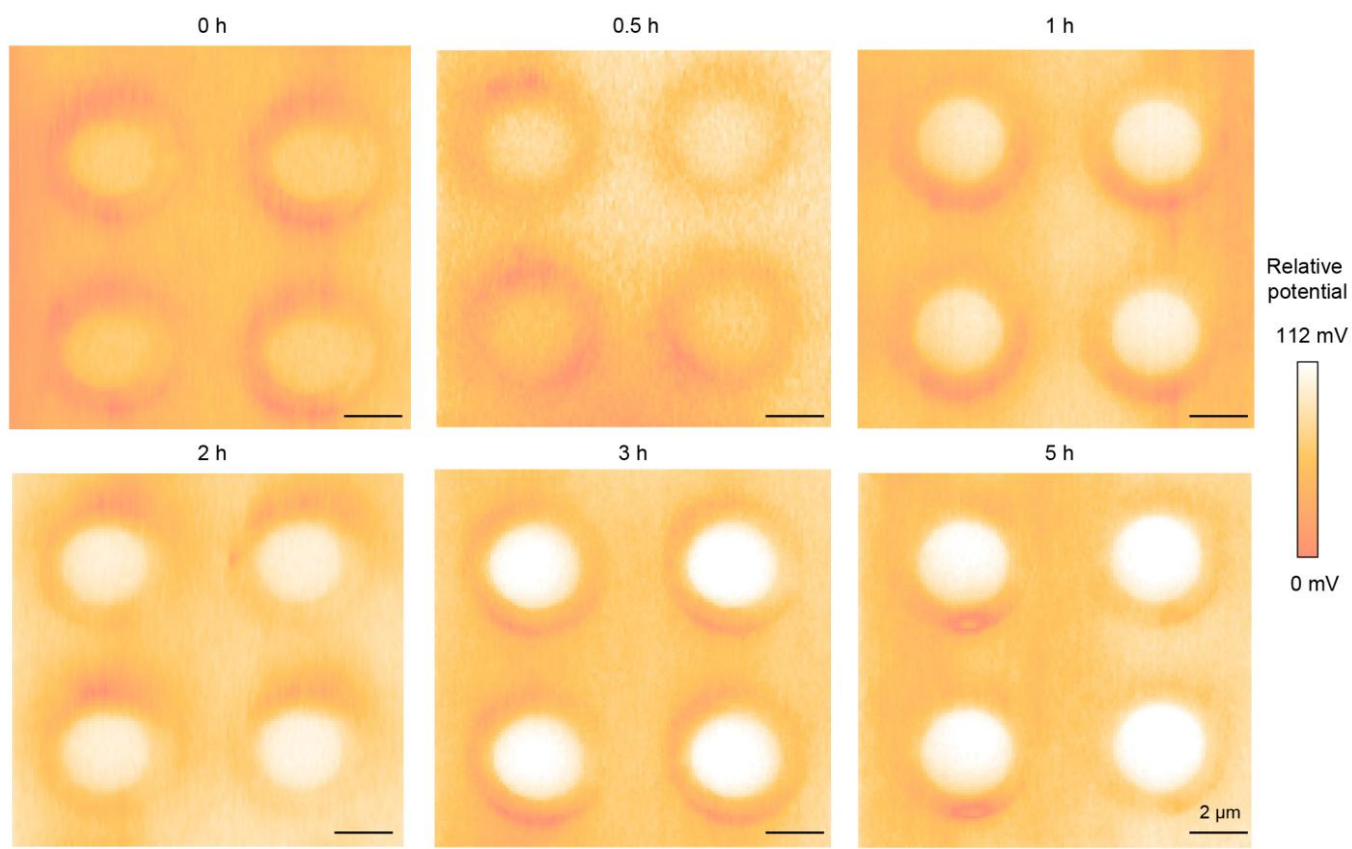

**Extended Data Fig. 3**

KPFM measurement of surface electrostatic potential in terms of template modified with different times. In the first 3 hours, the potential of modified Si (bottom of holes) was gradually increasing, which was saturated when the modification time continued to increase.

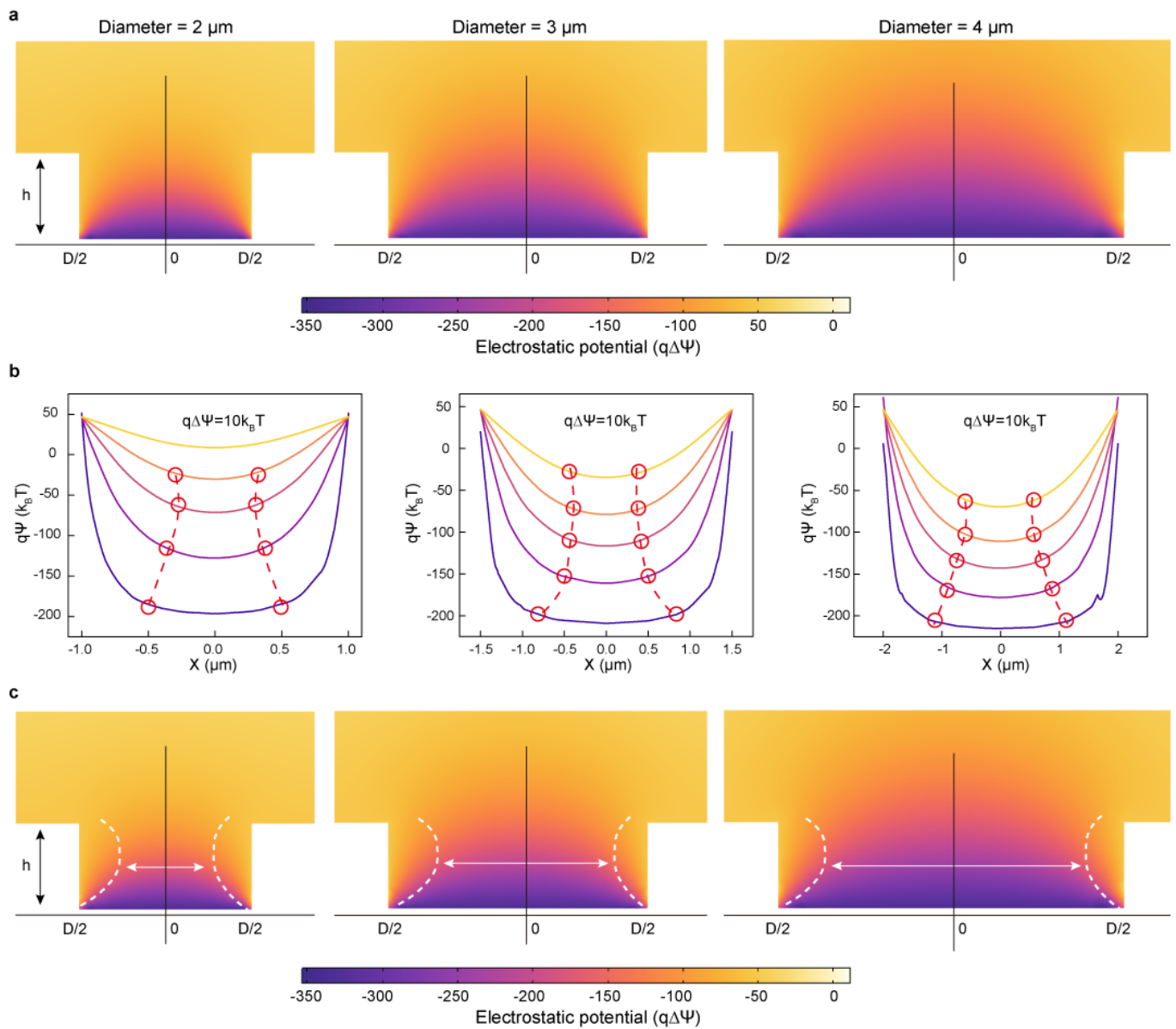

**Extended Data Fig. 4**

**a**, Distribution of calculated electrostatic potential in different-sized template holes. **b**, Extracted electrostatic potential curves at different depths of the hole, where the red circle represents the relative potential difference $q\Delta\psi$ is approximately 10 $k_BT$. The region inside the red circle is regarded as strong enough to efficiently trap NDs. **c**, The merged electrostatic potential and effective trapping region in different-sized template holes.

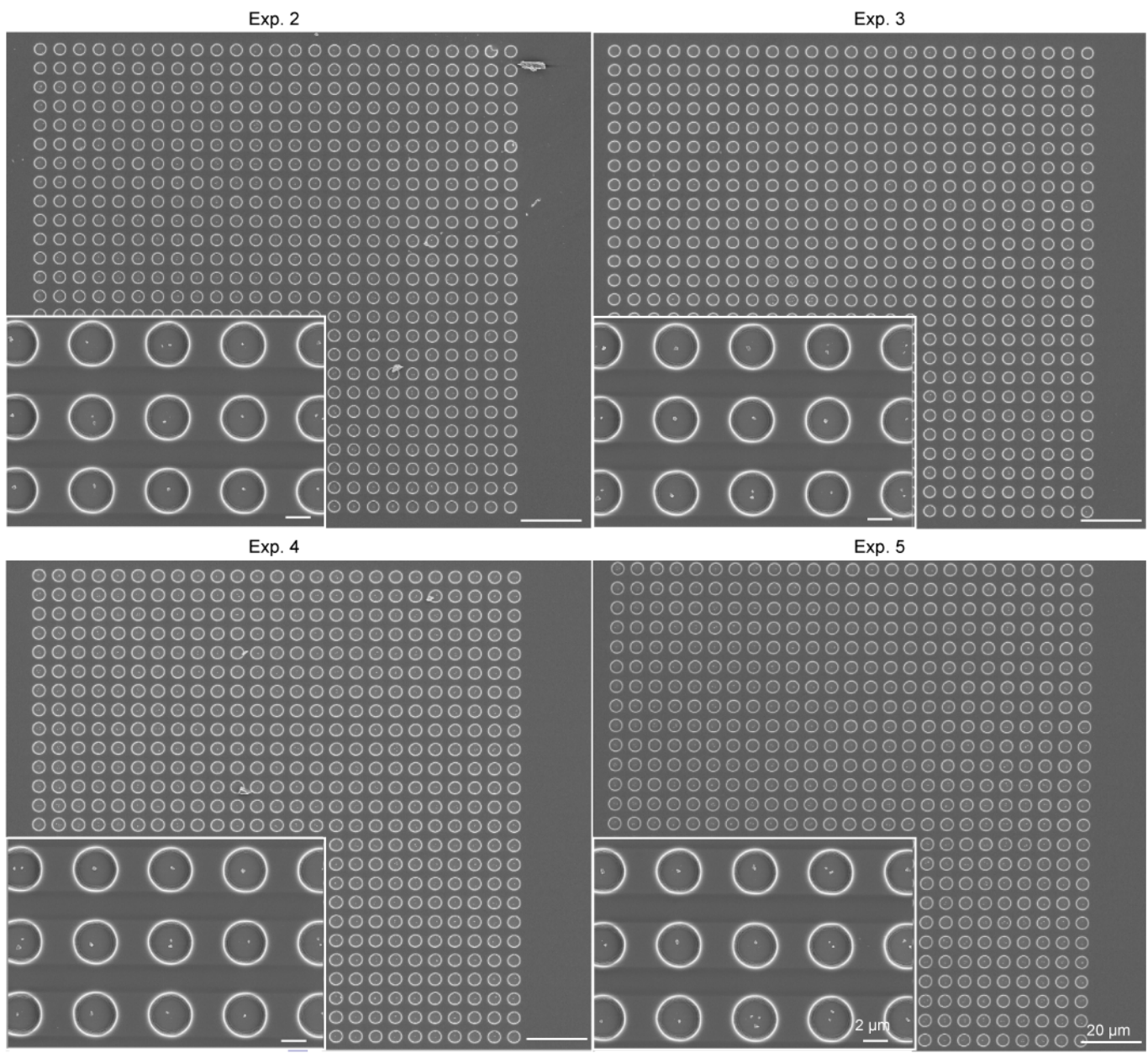

**Extended Data Fig. 5**

SEM images showing a 25 × 25 ND arrays before removing the template for four samples fabricated in different experiments.

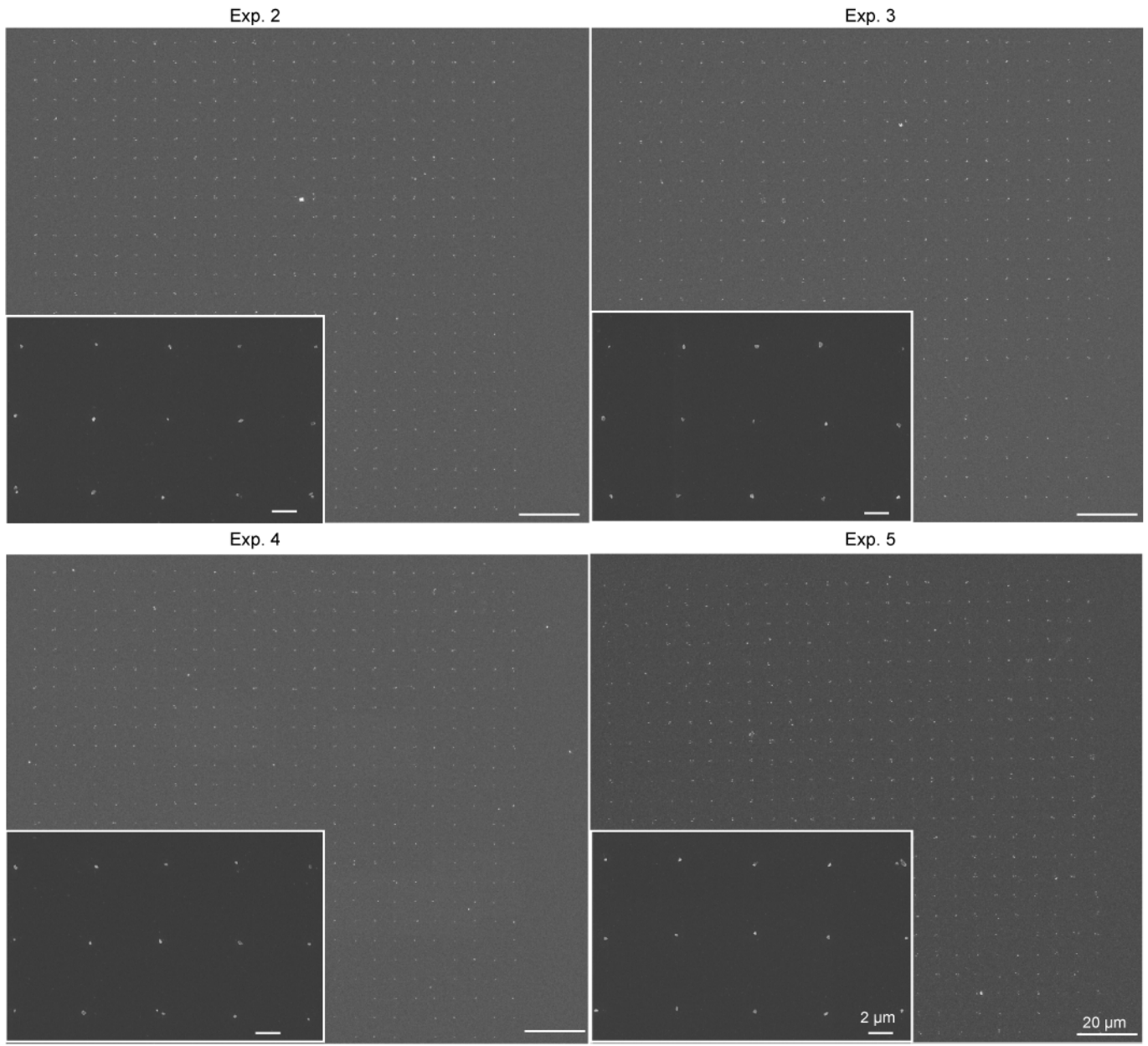

**Extended Data Fig. 6**

SEM images showing a 25 × 25 ND arrays after removing the template for four samples fabricated in different experiments.